# Polarizable-Vacuum (PV) representation of general relativity

H. E. Puthoff
*Institute for Advanced Studies at Austin*
*4030 W. Braker Lane, Suite 300, Austin, Texas 78759*
puthoff@aol.com

## ABSTRACT

Standard pedagogy treats topics in general relativity (GR) in terms of tensor formulations in curved space-time. Although mathematically straightforward, the curved space-time approach can seem abstruse to beginning students due to the degree of mathematical sophistication required. As a heuristic tool to provide insight into what is meant by a curved metric, we present a polarizable-vacuum (PV) representation of GR derived from a model by Dicke and related to the *"THεμ"* formalism used in comparative studies of gravitational theories.

## I. INTRODUCTION

Textbook presentations treat General Relativity (GR) in terms of tensor formulations in curved space-time. Such an approach captures in a concise and elegant way the interaction between masses, and their consequent motion. "Matter tells space how to curve, and space tells matter how to move [1]."

Although conceptually straightforward, the curved space-time approach can seem rather abstract to beginning students, and often lacking in intuitive appeal. During the course of development of GR over the years, however, alternative approaches have emerged that provide convenient methodologies for investigating metric changes in less abstract formalisms, and which yield heuristic insight into what is meant by a curved metric.

One approach that has a long history in GR studies, and that does have intuitive appeal, is what can be called the polarizable-vacuum (PV) representation of GR. Introduced by Wilson [2] and developed further by Dicke [3], the PV approach treats metric changes in terms of equivalent changes in the permittivity and permeability constants of the vacuum, $\varepsilon_o$ and $\mu_o$, essentially along the lines of the so-called *"THεμ"* methodology used in comparative studies of gravitational theories [4-6].

In brief, Maxwell's equations in curved space are treated in the isomorphism of a polarizable medium of variable refractive index in flat space [7]; the bending of a light ray near a massive body is modeled as due to an induced spatial variation in the refractive index of the vacuum near the body; the GR reduction in the velocity of light in a gravitational potential as compared to a flat-space reference frame at infinity is represented by an effective increase in the refractive index of the vacuum, and so forth. As elaborated in Refs. [3-7], PV modeling can be carried out in a self-consistent way so as to reproduce to appropriate order both the equations of GR, and the match to the classical experimental tests of those equations. In what follows we shall continually cross-reference PV-derived results to those obtained by conventional GR techniques to confirm that the PV approach as applied does not introduce spurious results.

## II. THE POLARIZABLE VACUUM

At this point one approach would be to introduce an appropriate Lagrangian for the PV formulation, and then derive equations of motion. Recognizing, however, that the PV approach is likely unfamiliar to most readers, we defer such formal derivation to Section IV. In its place we simply present and apply here certain key results of the more formal derivation to problems of interest, such as the three classical tests of GR. In this way we build up for the reader an intuitive picture of what is meant by a polarizable-vacuum representation of curved space-time.

Under non-curved-space conditions the electric flux vector $\boldsymbol{D}$ in a linear, homogeneous medium can be written

$$\boldsymbol{D} = \varepsilon \boldsymbol{E} = \varepsilon_o \boldsymbol{E} + \boldsymbol{P} \tag{1}$$

where $\varepsilon$ and $\varepsilon_o$ are the permittivities of the medium and of the vacuum, respectively, and the polarization $\boldsymbol{P}$ corresponds to the induced dipole moment per unit volume in the medium. Writing $\boldsymbol{P}$ in terms of the polarizability per unit volume, $\alpha_V$, as $\boldsymbol{P} = \alpha_V \boldsymbol{E}$, we obtain the symmetrical form

$$\boldsymbol{D} = \varepsilon_o \boldsymbol{E} + \alpha_V \boldsymbol{E} \, . \tag{2}$$

The above expression leads naturally to the interpretation of $\varepsilon_o$ as the polarizability per unit volume of the vacuum, treated as a medium in its own right. That this interpretation is correct is explicitly corroborated in detail by the picture of the vacuum as developed in quantum field theory. There it is shown that the vacuum acts as a polarizable medium by virtue of induced dipole moments resulting from the excitation of virtual electron-positron pairs [8].

With regard to representing curved-space conditions, the basic postulate for the PV approach is that the polarizability of the vacuum in the vicinity of, say, a mass differs from its asymptotic far-field value by virtue of vacuum polarization effects induced by the presence of the body. That is, we postulate for the vacuum itself

$$\boldsymbol{D} = \varepsilon\boldsymbol{E} = K \varepsilon_o \boldsymbol{E} \, , \tag{3}$$

where $K$ is the (altered) dielectric constant of the vacuum (typically a function of position) due to (GR-induced) vacuum polarizability changes under consideration. *Throughout the rest of our study the vacuum dielectric constant $K$ constitutes the key variable of interest.*

We next examine quantitatively the effects of a polarizable vacuum on the various measurement processes that form the basis of the PV approach to general relativity.

## A. Velocity of Light in a Vacuum of Variable Polarizability

We begin our discussion of the polarizable-vacuum model by examining constraints imposed by observation. An appropriate starting point is the expression for the fine structure constant,

$$\alpha = \frac{e^2}{4\pi\varepsilon_o \hbar c}, \quad \text{where} \quad c = \frac{1}{\sqrt{\mu_o \varepsilon_o}}. \tag{4}$$

By the conservation of charge for elementary particles, $e$ can be taken as a constant. Similarly, by conservation of angular momentum for a circularly polarized photon propagating through the vacuum (even with variable polarizability), $\hbar$ can be taken as a constant. Given that $\varepsilon_o$ and $c$ can be expected with a variable vacuum polarizability to change to $\varepsilon(K) = K\varepsilon_o$ and $c'(K) = 1/\sqrt{\mu(K)\varepsilon(K)}$, the fine structure constant takes the form

$$\alpha = \frac{e^2}{4\pi\varepsilon_o \hbar c}\sqrt{\frac{\mu(K)/\mu_o}{K}} \tag{5}$$

which is potentially a function of $K$.

Studies that consider the possibility of the variability of fundamental constants under varying cosmological conditions, however, require that the fine structure constant remain constant in order to satisfy the constraints of observational data [9-11]. Under this constraint we obtain from Eq. (5) $\mu(K) = K\mu_o$; thus the permittivity and permeability constants of the vacuum must change together with vacuum polarizability as

$$\varepsilon_o \rightarrow \varepsilon = K\varepsilon_o, \quad \mu_o \rightarrow \mu = K\mu_o. \tag{6}$$

This transformation, which maintains constant the ratio $\sqrt{\mu/\varepsilon} = \sqrt{\mu_o/\varepsilon_o}$ (the impedance of free space) is just what is required to maintain electric-to-magnetic energy ratios constant during adiabatic movement of atoms from one point to another of differing vacuum polarizability [3]. Detailed analysis shows that it is also a necessary condition in the $TH\varepsilon\mu$ formalism for an electromagnetic test body to fall in a gravitational field with a composition-independent acceleration (WEP, or weak equivalence principle, verified by Eötvös-type experiments) [4-6]. Finally, this condition must be satisfied by any metric theory of gravity, which constitutes the class of viable gravity theories.

The above arguments therefore lead us to our first major conclusion concerning the effects of a variable vacuum polarizability; namely, the permittivity and permeability constants, $\varepsilon$ and $\mu$, change linearly with the vacuum dielectric constant $K$ as given in Eq. (6), and the velocity of light therefore changes inversely with $K$ in accordance with

$$c \,(= 1/\sqrt{\mu_o \varepsilon_o}) \longrightarrow c/K \,(= 1/\sqrt{\mu\varepsilon}). \tag{7}$$

Thus, the dielectric constant of the vacuum plays the role of a variable refractive index under conditions in which vacuum polarizability is assumed to change in response to GR-type influences.

### B. Energy in a Vacuum of Variable Polarizability

As will be made explicit in the detailed derivations later, the PV treatment of GR effects is based on the use of equations that hold in special relativity, but with the modification that the velocity of light $c$ in the Lorentz transformations and elsewhere is replaced by the velocity of light in a medium of variable refractive index, $c/K$, as given in the previous section. Expressions such as $E = mc^2$ are still valid, but now take into account that $c \to c/K$, and $E$ and $m$ may be functions of $K$, and so forth.

With regard to expressions for energy, Dicke has shown by application of a limited principle of equivalence that the energy of a system whose value is $E_O$ in flat space $(K = 1)$ takes on the value

$$E = \frac{E_O}{\sqrt{K}} \qquad (8)$$

in a region where $K \neq 1$ [3]. This is due to that fact that the self-energy of a system changes in response to changes in the local vacuum polarizability. This is analogous to the change in the stored energy of a charged air capacitor during transport to a region of differing dielectric constant.

The energy relationship given by Eq. (8) also implies, via $E_O = m_O c^2$, which becomes $E(K) = m(K)(c/K)^2$, a corollary change in mass

$$m = m_O K^{3/2} \quad , \qquad (9)$$

again a consequence of the change in self-energy.

### C. Rod and Clock (Metric) Changes in a Vacuum of Variable Polarizability

Another consequence of the change in energy as a function of vacuum polarizability is a change in associated frequency processes which, by the quantum condition $E = \hbar \omega$ and Eq. (8), takes the form

$$\omega = \frac{\omega_O}{\sqrt{K}} \quad . \qquad (10)$$

This, as we shall see, is responsible for the red shift in light emitted from an atom located in a gravitational potential.

From the reciprocal of Eq. (10) we find that time intervals marked by such processes are related by

$$\Delta t = \Delta t_O \sqrt{K} \ . \tag{11}$$

Therefore, in a gravitational potential (where it will be shown that $K > 1$) the time interval between clock ticks is increased (that is, the clock runs slower) relative to a reference clock at infinity.

With regard to effects on measuring rods, we note that, for example, the radius of the ground-state Bohr orbit of a hydrogen atom

$$\Delta r_O = \frac{\hbar}{m_O c} \frac{1}{\alpha} \tag{12}$$

becomes (with $c \rightarrow c/K$, $m_O \rightarrow m$, and $\alpha$ constant as discussed earlier)

$$\Delta r = \frac{\Delta r_O}{\sqrt{K}} \ . \tag{13}$$

Other measures of length such as the classical electron radius or the Compton wavelength of a particle lead to the relationship Eq. (13) as well, so this relationship is general. This dependence of fundamental length measures on the variable $K$ indicates that the dimensions of material objects adjust in accordance with local changes in vacuum polarizability - thus there is no such thing as a perfectly rigid rod. From the standpoint of the PV approach this is the genesis of the variable metric that is of such significance in GR studies.

We are now in a position to define precisely what is meant by the label "curved space." In the vicinity of, say, a planet or star, where $K > 1$, if one were to take a ruler and measure along a radius vector $R$ to some circular orbit, and then measure the circumference $C$ of that orbit, one would obtain $C < 2\pi R$ (as for a concave curved surface). This is a consequence of the ruler being relatively shorter during the radial measuring process (see Eq. (13)) when closer to the body where $K$ is relatively greater, as compared to its length during the circumferential measuring process when further from the body. Such an influence on the measuring process due to induced polarizability changes in the vacuum near the body leads to the GR concept that the presence of the body "influences the metric," and correctly so.

Of special interest is the measurement of the velocity of light with "natural" (i.e., physical) rods and clocks in a gravitational potential which have become "distorted" in accordance with Eqns. (11) and (12). Let the end points of a measurement be a "true" distance $\Delta X$ apart, and let the time for propagation of a light signal from one end point to the other be a "true" time interval $\Delta T$. Since the measurement is assumed to take place in a gravitational potential with a propagation velocity $c/K$, $\Delta X$ and $\Delta T$ are related by

$$\frac{\Delta X}{\Delta T} = \frac{c}{K} \ . \tag{14}$$

However, a rod whose length is $\Delta r_O$ at infinity will have shortened in the gravitational potential in accordance with Eq. (13), and therefore the *measured* length $\Delta X_m$ will be given not by $\Delta X$ but by $\Delta X_m = \Delta X \sqrt{K}$. Similarly, a clock whose rate is determined by Eq. (11), and therefore runs slower in the gravitational potential, will yield for a *measured* propagation time $\Delta T_m = \Delta T / \sqrt{K}$. As a result the *measured* velocity of light, $v_m$, is given by

$$v_m = \frac{\Delta X_m}{\Delta T_m} = \frac{\Delta X \sqrt{K}}{\Delta T / \sqrt{K}} = \frac{c}{K} \bullet K = c. \tag{15}$$

We thus arrive at the interesting (and significant) result that the measured velocity of light obtained by the use of physical rods and clocks renormalizes from its "true" value $c/K$ to the value $c$. The PV formalism therefore maintains the universal constancy of the locally *measured* velocity of light.

### D. The Metric Tensor

At this point we can make a crossover connection to the standard metric tensor concept that characterizes the conventional GR formulation. In flat space a (4-dimensional) infinitesimal interval is given by the expression

$$ds^2 = c^2 dt_O^2 - \left(dx_O^2 + dy_O^2 + dz_O^2\right). \tag{16}$$

If rods were rigid and clocks non-varying in their performance in regions of differing vacuum polarizability, then the above expression would hold universally. However, a $dx_O$-length measuring rod placed in a region where $K > 1$, for example, shrinks according to Eq. (13) to $dx = dx_O / \sqrt{K}$. Therefore, the infinitesimal length which would measure $dx_O$ were the rod to remain rigid is now expressed in terms of the $dx$-length rod as $dx_O = \sqrt{K} dx$. With a similar argument based on Eq. (11) holding for clock rate, Eq. (16) can be written

$$ds^2 = \frac{1}{K} c^2 dt^2 - K(dx^2 + dy^2 + dz^2). \tag{17}$$

Therefore, the infinitesimal interval takes on the form

$$ds^2 = g_{ij} dx^i dx^j, \tag{18}$$

where $g_{ij}$ in the above expression defines the metric tensor, and

$$dx^0 = c\,dt, \; g_{00} = 1/K, \; g_{11} = g_{22} = g_{33} = -K, \; g_{ij} = 0 \; for \; i \neq j. \tag{19}$$

The metric tensor in this form defines an *isotropic coordinate system*, familiar in GR studies.

# III. CLASSICAL EXPERIMENTAL TESTS OF GENERAL RELATIVITY IN THE PV MODEL

In the previous sections we have established the concept of the polarizable vacuum and the effects of polarizability changes on metric (rods and clocks) behavior. In particular, we found that metric changes can be specified in terms of a single parameter $K$, the dielectric constant of the vacuum. This is the basis of the PV approach to GR.

In this section we shall explore, with the aid of expressions to be derived in detail in Section IV, specifically how $K$ changes in the presence of mass, and the effects generated thereby. The effects of major interest at this point comprise the three classical tests of GR; namely, the gravitational redshift, the bending of light and the advance of the perihelion of Mercury. These examples constitute a good testbed for demonstrating the techniques of the PV alternative to the conventional GR curved-space approach.

For the spherically symmetric mass distribution of a star or planet it will be shown later from the basic postulates of the PV approach that the appropriate PV expression for the vacuum dielectric constant $K$ is given by the exponential form

$$K = e^{2GM/rc^2} , \qquad (20)$$

where $G$ is the gravitational constant, $M$ is the mass, and $r$ is the distance from the origin located at the center of the mass $M$. For comparison with expressions derived by conventional GR techniques, it is sufficient for our purposes to restrict consideration to a weak-field approximation based on expansion of the exponential to second order, viz.,

$$K \approx 1 + \frac{2GM}{rc^2} + \frac{1}{2}\left(\frac{2GM}{rc^2}\right)^2 . \qquad (21)$$

## A. Gravitational Redshift

In a gravitation-free part of space, photon emission from an excited atom takes place at some frequency $\omega_O$, uninfluenced by vacuum polarizability changes. That same emission process taking place in a gravitational field, however, will, according to Eq. (10), have its emission frequency altered (redshifted) to $\omega = \omega_O/\sqrt{K}$. With the first-order correction to $K = 1$ given by the first two terms in Eq. (21), emission by an atom located on the surface of a body of mass $M$ and radius $R$ will therefore experience a redshift by an amount

$$\frac{\Delta\omega}{\omega_O} = \frac{\omega - \omega_O}{\omega_O} \approx -\frac{GM}{Rc^2} , \qquad (22)$$

where we take $GM/Rc^2 \ll 1$. Once emitted, the frequency of the photon remains constant during its propagation to a relatively gravitation-free part of space where its frequency can then be compared against that of local emission, and the spectral shift given by Eq. (22) observed.

Measurement of the redshift of the sodium $D_1$ line emitted on the surface of the sun and received on earth has verified Eq. (22) to a precision of 5% [12].

Experiments carried out on the surface of the earth involving the comparison of photon frequencies at different heights have improved the accuracy of verification still further to a precision of 1% [13-14]. In this case, with the two ends of the experiment separated by a vertical height $h$, the first-order frequency shift is calculated with the aid of Eqns. (10) and (21) as

$$\frac{\Delta\omega}{\omega} \approx \frac{GM}{R^2 c^2} h, \qquad (23)$$

where $M$ and $R$ are the mass and radius of the earth. This experiment, carried out at Harvard University, required a measurement accuracy of $\Delta\omega/\omega \sim 10^{-15}$ for a height $h = 22.5$ meters. It was accomplished by the use of Mössbauer-effect measurement of the difference between $\gamma$-ray emission and absorption frequencies at the two ends of the experiment.

### B. Bending of Light Rays

We consider now the case of a light ray passing in close proximity to a massive body, say, the sun. Given that the velocity of light in a gravitational potential is given by $c/K$, and given that $K$ varies as shown in Eq. (21), the light ray in essence passes through a medium of variable refractive index.

Since $K$ rises in value near the body, there is a corresponding reduction in the velocity of light in that region. Therefore, for a light ray grazing the body, that part of the wavefront closest to the body will, by virtue of its reduced velocity, cause the wavefront to "wheel about" the body, so to speak. As a result the light ray is deflected toward the body. This deflection, in GR terms, yields a measure of "local curvature," while in the PV approach it is interpreted as a measure of the spatially dependent vacuum polarizability.

Quantitatively, the magnitude of the deflection can be calculated as follows, where we use the sun as an example. Since the deflection is small (less than two seconds of arc), small-angle approximations can be used throughout.

As a light ray grazes the sun, the wavefront tilt, and hence the deflection angle, is found by calculating the accumulated difference in wavefront advancement across the ray (transverse to the direction of propagation) due to the slight difference in phase velocity. Since the velocity of light is given by $v_L = c/K$, to first order Eq. (21) yields

$$v_L = c/K \approx c \bigg/ \left(1 + \frac{2GM}{rc^2}\right) \approx c\left(1 - \frac{2GM}{rc^2}\right). \qquad (24)$$

With the geometry of a grazing ray as shown in Fig. 1(a), Eq. (24) can be written

$$v_L \approx c\left[1 - \frac{2GM}{c^2}\frac{1}{\sqrt{(R+\delta)^2 + z^2}}\right] . \tag{25}$$

Therefore, the differential velocity across the ray is (for $\delta \ll R$)

$$\Delta v_L \approx \frac{2GM}{c}\frac{R\delta}{(R^2 + z^2)^{3/2}} . \tag{26}$$

Reference to Fig. 1(b) indicates that, as the ray travels a distance $dz \approx c\, dt$, the differential velocity across the ray results in an accumulated wavefront difference $\Delta z$ given by

$$\Delta z = \Delta v_L dt \approx \frac{2GM}{c^2}\frac{R\delta}{(R^2 + z^2)^{3/2}} dz \tag{27}$$

and an accumulated tilt angle

$$d\alpha \approx \tan(d\alpha) = \frac{\Delta z}{\delta} \approx \frac{2GM}{c^2}\frac{R}{(R^2 + z^2)^{3/2}} dz . \tag{28}$$

Integrating over the entire ray path from $z = -\infty$ to $z = +\infty$ we obtain

$$\alpha \approx \frac{4GM}{Rc^2} . \tag{29}$$

For the sun this gives 1.75 seconds of arc, a value first verified within experimental error by measurements of grazing starlight images made during total eclipses of the sun in 1919. Further confirmation was obtained in the form of deflection by the planet Jupiter of waves from a strong celestial radio source by an amount of 300 microseconds of arc [15].

### C. Advance of the Perihelion of Mercury

The third classical test of GR theory involves a prediction of a slight precession (rotation) of the elliptical orbit of a planet or satellite with each orbital revolution. For even the closest planet to the sun, Mercury, however, the advance of the perihelion amounts to only 43 seconds of arc per century, and is therefore an extremely small effect.

Analysis of the perihelion advance in the PV approach exhibits quite explicitly how changes in vacuum properties lead to GR effects. We begin with the standard Lagrangian for a *free* particle [16],

$$L = -mc^2\sqrt{1 - \left(\frac{v}{c}\right)^2} , \tag{30}$$

only we note that the mass $m(K)$, given by Eq. (9), and the velocity of light $c(K)$, given by Eq. (7), are now functions of the vacuum dielectric constant $K$,

$$L = -m(K)c(K)^2\sqrt{1-\left(\frac{v}{c/K}\right)^2} = -m_o K^{3/2}\left(\frac{c}{K}\right)^2\sqrt{1-\frac{\left(\dot{r}^2+r^2\dot{\theta}^2\right)}{(c/K)^2}}. \tag{31}$$

With $K$ given by Eq. (20), the above takes the form

$$L = -m_o c^2 e^{-GM/rc^2}\sqrt{1-\frac{\left(\dot{r}^2+r^2\dot{\theta}^2\right)e^{4GM/rc^2}}{c^2}}. \tag{32}$$

Substitution into Lagrange's general equations of motion

$$\frac{d}{dt}\left(\frac{\partial L}{\partial \dot{q}}\right) - \frac{\partial L}{\partial q} = 0 \tag{33}$$

then leads to the specific equations of motion

$$\frac{d}{dt}\left(mr^2\dot{\theta}\right) = 0, \tag{34a}$$

$$\frac{d}{dt}\left(m\dot{r}\right) = -\frac{GMm}{r^2}e^{-4GM/rc^2} + mr\dot{\theta}^2 - \frac{2GMm}{r^2c^2}\left(\dot{r}^2+r^2\dot{\theta}^2\right), \tag{34b}$$

where $m = m_o K^{3/2} = m_o \times \exp(3GM/rc^2)$ and, once derivatives are taken, we assume $\sqrt{1-(v/c(K))^2} \approx 1$ for nonrelativistic motion.

The solution to Eq. (34a) yields an expression for the angular momentum $\ell$,

$$mr^2\dot{\theta} = \ell. \tag{35}$$

Substitution into Eq. (34b), followed by elimination of the variable $t$ by means of $d/dt = \dot{\theta}\, d/d\theta$, then leads to an equation for the planetary trajectory in the $(r, \theta)$ plane,

$$\frac{d}{d\theta}\left(\frac{1}{r^2}\frac{dr}{d\theta}\right) = -\frac{GMm_o^2}{\ell^2}e^{2GM/rc^2} + \frac{1}{r} - \frac{2GM}{c^2}\left[\frac{1}{r^2} + \frac{1}{r^4}\left(\frac{dr}{d\theta}\right)^2\right]. \tag{36}$$

With a change of variable to $u = 1/r$, we obtain, finally,

$$\frac{d^2u}{d\theta^2} + u = \frac{GMm_O^2}{\ell^2} e^{2GMu/c^2} + \frac{2GM}{c^2}\left[u^2 + \left(\frac{du}{d\theta}\right)^2\right] . \tag{37}$$

Taking $exp(2GM/rc^2) \approx 1 + 2GM/rc^2$, that is, $exp(2GMu/c^2) \approx 1 + 2GMu/c^2$, we assume a solution of the form

$$u = A(1 + e\cos\gamma\theta) , \tag{38}$$

where here $e$ is the eccentricity of the orbit ($e = 0.206$ for Mercury). This leads to good approximation to an expression for $\gamma$ of the form

$$\gamma = \sqrt{1 - 6\left(\frac{GMm_O}{\ell c}\right)^2} \approx 1 - 3\left(\frac{GMm_O}{\ell c}\right)^2 . \tag{39}$$

For one revolution of the planet in its solar orbit, the second term in Eq. (39) represents a correction to the nonprecessing Newtonian orbit. In particular, it corresponds to a perihelion advance per orbit of

$$\Delta\theta = 2\pi \times 3\left(\frac{GMm_O}{\ell c}\right)^2 = 6\pi\left(\frac{GMm_O}{\ell c}\right)^2 , \tag{40}$$

which for the planet Mercury accumulates to the observed value of 43 seconds of arc per century.

From the standpoint of conventional GR theory the predicted and observed perihelion advance constitutes remarkable corroboration of the curved-space concept. From the standpoint of the PV approach such a result can be interpreted as corroboration of the fact that the dielectric constant of the vacuum in the vicinity of mass is modified by the presence of that mass. The epistemological link between the two approaches can be seen in the fact that in neither is gravity treated as a force *per se*. Rather, in both cases the test body is treated as a *free* particle (note the use of a free-particle Lagrangian in the PV approach) whose trajectory is determined "by the metric properties of space itself."

### D. Comparison of the Metric Tensors in the GR and PV Approaches

Having shown by specific calculation that the PV approach to the three classical tests of GR reproduces the traditional GR results, we now establish the underlying generality of such correspondence.

In standard textbook treatments of the three classical tests of GR one begins with the Schwarzschild metric, which in isotropic coordinates is written [17]

$$ds^2 = g_{ij} dx^i dx^j$$

$$= \left( \frac{1 - \frac{GM}{2rc^2}}{1 + \frac{GM}{2rc^2}} \right)^2 c^2 dt^2 - \left(1 + \frac{GM}{2rc^2}\right)^4 \left(dr^2 + r^2 d\theta^2 + r^2 \sin^2\theta d\phi^2 \right) . \tag{41}$$

Expanding the metric tensor for small departures from flatness as a Maclaurin series in $(GM/rc^2)$, we obtain

$$g_{00} = \left( \frac{1 - \frac{GM}{2rc^2}}{1 + \frac{GM}{2rc^2}} \right)^2 = 1 - 2\left(\frac{GM}{rc^2}\right) + 2\left(\frac{GM}{rc^2}\right)^2 - ... \tag{42}$$

$$g_{11} = g_{22} = g_{33} = -\left(1 + \frac{GM}{2rc^2}\right)^4 = -\left[1 + 2\left(\frac{GM}{rc^2}\right) + ...\right] . \tag{43}$$

Similarly, in the PV approach one begins with the exponential metric defined by Eqns. (17) - (20),

$$ds^2 = g_{ij} dx^i dx^j$$
$$= e^{-2GM/rc^2} c^2 dt^2 - e^{2GM/rc^2} \left(dr^2 + r^2 d\theta^2 + r^2 \sin^2\theta d\phi^2 \right) . \tag{44}$$

This, when expanded to the same order as the Schwarzschild metric tensor above for small departures from unity vacuum dielectric constant, yields

$$g_{00} = e^{-2GM/rc^2} = 1 - 2\left(\frac{GM}{rc^2}\right) + 2\left(\frac{GM}{rc^2}\right)^2 - ... , \tag{45}$$

$$g_{11} = g_{22} = g_{33} = -e^{2GM/rc^2} = -\left[1 + 2\left(\frac{GM}{rc^2}\right) + ...\right] . \tag{46}$$

Comparison of Eqns. (45) - (46) with Eqns. (42) - (43) reveals that, to the order of expansion shown, the two metric tensors are identical. Since the three classical tests of GR do not require terms beyond these explicitly displayed, the agreement between theory and experiment is accounted for equally in both the conventional GR and in the alternative PV formalisms [18].

## IV. COUPLED MATTER-FIELD EQUATIONS

In the preceding section we have seen that the three classical tests of GR theory, redshift, bending of light rays, and the advance of the perihelion of Mercury, are accounted for in the PV formalism on the basis of a variable vacuum dielectric constant, $K$. To carry that out we stated without proof that the appropriate mathematical form for the variation in $K$ induced by the presence of mass is an exponential form, which, under the weak-field conditions prevailing in the solar system, can be treated in terms of the series expansion

$$K = e^{2GM/rc^2} = 1 + 2\left(\frac{GM}{rc^2}\right) + \ldots \quad . \tag{47}$$

In this section we show how the exponential form is derived from first principles, and, in the process, establish the general approach to the derivation of field equations as well as the equations for particle motion. The approach consists of following standard Lagrangian techniques as outlined, for example, in Ref. 16, but with the proviso that in our case the dielectric constant of the vacuum is treated as a variable function of time and space.

### A. Lagrangian Approach

As stated earlier, the standard Lagrangian for a free particle is given by the form Eq. (30)

$$L^p = -mc^2\sqrt{1-\left(\frac{v}{c}\right)^2} \quad , \tag{48}$$

which, in the presence of a variable vacuum dielectric constant $K$ is modified to Eq. (31) with the aid of Eqns. (7) and (9) to read

$$L^p = -\frac{m_O c^2}{\sqrt{K}}\sqrt{1-\left(\frac{v}{c/K}\right)^2} \quad . \tag{49}$$

This implies a Lagrangian *density* for the particle of

$$L^p_d = -\frac{m_O c^2}{\sqrt{K}}\sqrt{1-\left(\frac{v}{c/K}\right)^2}\,\delta^3(\mathbf{r}-\bar{\mathbf{r}}) \quad , \tag{50}$$

where $\delta^3(\mathbf{r}-\bar{\mathbf{r}})$ is the three-dimensional delta function that serves to locate the (point) particle at $\mathbf{r} = \bar{\mathbf{r}}$.

Following standard procedure, the particle Lagrangian density can be extended to the case of interaction with electromagnetic fields by the addition of terms involving the scalar and vector potentials $(\Phi, \mathbf{A})$,

$$L_d^p = -\left(\frac{m_O c^2}{\sqrt{K}}\sqrt{1-\left(\frac{v}{c/K}\right)^2} + q\Phi - q\mathbf{A}\cdot\mathbf{v}\right)\delta^3(\mathbf{r}-\bar{\mathbf{r}}), \tag{51}$$

where $(\Phi, \mathbf{A})$ are related to the electromagnetic field vectors $(\mathbf{E}, \mathbf{B})$ by

$$\mathbf{E} = -\nabla\Phi - \frac{\partial \mathbf{A}}{\partial t}, \quad \mathbf{B} = \nabla \times \mathbf{A}. \tag{52}$$

The Lagrangian density for the electromagnetic fields themselves, as in the case of the particle Lagrangian, is given by the standard expression (see, e.g., Ref. 16), except that again $K$ is treated as a variable,

$$L_d^{em} = -\frac{1}{2}\left(\frac{B^2}{K\mu_O} - K\varepsilon_O E^2\right). \tag{53}$$

We now need a Lagrangian density for the dielectric constant variable $K$, which, being treated as a scalar variable, must take on the standard Lorentz-invariant form for propagational disturbances of a scalar,

$$L_d^K = -\lambda f(K)\left[(\nabla K)^2 - \frac{1}{(c/K)^2}\left(\frac{\partial K}{\partial t}\right)^2\right], \tag{54}$$

where $f(K)$ is an arbitrary function of $K$. As indicated by Dicke in the second citation of Ref. 3, a correct match to experiment requires that we take $\lambda = c^4/32\pi G$ and $f(K) = 1/K^2$; thus,

$$L_d^K = -\frac{\lambda}{K^2}\left[(\nabla K)^2 - \frac{1}{(c/K)^2}\left(\frac{\partial K}{\partial t}\right)^2\right]. \tag{55}$$

We can now write down the total Lagrangian density for matter-field interactions in a vacuum of variable dielectric constant,

$$L_d = -\left(\frac{m_O c^2}{\sqrt{K}}\sqrt{1-\left(\frac{v}{c/K}\right)^2} + q\Phi - q\mathbf{A}\cdot\mathbf{v}\right)\delta^3(\mathbf{r}-\bar{\mathbf{r}}) - \frac{1}{2}\left(\frac{B^2}{K\mu_O} - K\varepsilon_O E^2\right)$$
$$-\frac{\lambda}{K^2}\left[(\nabla K)^2 - \frac{1}{(c/K)^2}\left(\frac{\partial K}{\partial t}\right)^2\right]. \tag{56}$$

## B. General Matter-Field Equations

Variation of the Lagrangian density $\delta \int L_d dxdydzdt$ with regard to the particle variables as per standard action principle techniques (Ref. 16) leads to our first important equation, the equation for particle motion in a variable dielectric vacuum,

$$\frac{d}{dt}\left[\frac{(m_O K^{3/2})\mathbf{v}}{\sqrt{1-\left(\frac{v}{c/K}\right)^2}}\right] = q(\mathbf{E} + \mathbf{v}\times\mathbf{B}) - \nabla\left(\frac{m_O c^2}{\sqrt{K}}\sqrt{1-\left(\frac{v}{c/K}\right)^2}\right) \quad (57)$$

or

$$\frac{d}{dt}\left[\frac{(m_O K^{3/2})\mathbf{v}}{\sqrt{1-\left(\frac{v}{c/K}\right)^2}}\right] = q(\mathbf{E} + \mathbf{v}\times\mathbf{B}) + \frac{(m_O c^2/\sqrt{K})}{\sqrt{1-\left(\frac{v}{c/K}\right)^2}}\left[\frac{1+\left(\frac{v}{c/K}\right)^2}{2}\right]\frac{\nabla K}{K}. \quad (58)$$

We see that accompanying the usual Lorentz force is an additional dielectric force proportional to the gradient of the vacuum dielectric constant. This term is equally effective with regard to both charged and neutral particles and accounts for the familiar gravitational potential, whether Newtonian in form or, as in the last section, taken to higher order to account for GR effects. One note of passing interest is the fact that as $m_O \to 0$, but $v \to c/K$, the deflection of a zero-mass particle (e.g., a photon) in a gravitational field is twice that of a slow-moving particle ($v \to 0$), an important result in GR dynamics.

Variation of the Lagrangian density with regard to the $K$ variable leads to the second equation of interest, an equation for the generation of GR vacuum polarization effects due to the presence of matter and fields. (In the final expression we use $\nabla^2 K = (1/2K)(\nabla K)^2 + 2\sqrt{K}\nabla^2\sqrt{K}$ to obtain a form convenient for the following discussion.)

$$\nabla^2\sqrt{K} - \frac{1}{(c/K)^2}\frac{\partial^2\sqrt{K}}{\partial t^2} = -\frac{\sqrt{K}}{4\lambda}\left\{\frac{(m_O c^2/\sqrt{K})}{\sqrt{1-\left(\frac{v}{c/K}\right)^2}}\left[\frac{1+\left(\frac{v}{c/K}\right)^2}{2}\right]\delta^3(\mathbf{r}-\bar{\mathbf{r}})\right.$$

$$\left. + \frac{1}{2}\left(\frac{B^2}{K\mu_O} + K\varepsilon_O E^2\right) - \frac{\lambda}{K^2}\left[(\nabla K)^2 + \frac{1}{(c/K)^2}\left(\frac{\partial K}{\partial t}\right)^2\right]\right\}. \quad (59)$$

Thus we see that changes in the vacuum dielectric constant *K* are driven by mass density (first term), EM energy density (second term), and the vacuum polarization energy density itself (third term). The fact that the latter term corresponds to the energy density of the *K* variable can be seen by the following argument. We start with the Lagrangian density Eq. (55), define the momentum density as usual by $\pi = \partial L_d^K / \partial(\partial K/\partial t)$, and form the Hamiltonian energy density by standard techniques to obtain

$$H_d^K = \pi\left(\frac{\partial K}{\partial t}\right) - L_d^K = \frac{\lambda}{K^2}\left[(\nabla K)^2 + \frac{1}{(c/K)^2}\left(\frac{\partial K}{\partial t}\right)^2\right]. \tag{60}$$

Again, in passing, of interest is the fact that the energy densities of the EM fields and the *K* variable enter into Eq. (59) with opposite signs. Therefore, EM field effects can counteract gravitational field effects. This will become more apparent when we examine the so-called "electrogravitic repulsion forces" associated with the Reissner-Nordstrøm solution to Eq. (59).

Eqns. (58) and (59), together with Maxwell's equations for propagation in a medium with variable dielectric constant, thus constitute the master equations to be used in discussing matter-field interactions in a vacuum of variable dielectric constant as required in the PV formulation of GR.

## V. STATIC FIELD SOLUTIONS

We bring our tutorial on the PV approach to a close by demonstrating application of the field Eq. (59) to two static field cases with spherical symmetry: derivation of the expression introduced earlier for the gravitational field alone, and derivation of the corresponding expression for charged masses.

### A. Static Fields (Gravitational)

In space surrounding an uncharged spherical mass distribution the static solution *(∂K/∂t = 0)* to Eq. (59) is found by solving

$$\nabla^2 \sqrt{K} = \frac{1}{4K^{3/2}}(\nabla K)^2 = \frac{1}{\sqrt{K}}\left(\nabla\sqrt{K}\right)^2 \tag{61a}$$

or

$$\frac{d^2\sqrt{K}}{dr^2} + \frac{2}{r}\frac{d\sqrt{K}}{dr} = \frac{1}{\sqrt{K}}\left(\frac{d\sqrt{K}}{dr}\right)^2, \tag{61b}$$

where we have used $(\nabla K)^2 = 4K\left(\nabla\sqrt{K}\right)^2$.

The solution that satisfies the Newtonian limit is given by

$$\sqrt{K} = e^{GM/rc^2} \tag{62a}$$

or

$$K = e^{2GM/rc^2} = 1 + 2\left(\frac{GM}{rc^2}\right) + \ldots \;, \tag{62b}$$

which can be verified by substitution into the equation for particle motion, Eq. (58). We have thus derived from first principles the exponential form of the variable dielectric constant in the vicinity of a mass as used in earlier sections. As indicated in Section III-D, this solution reproduces to appropriate order the standard GR Schwarzschild metric predictions as they apply to the weak-field conditions prevailing in the solar system.

## B. Static Fields (Gravitational Plus Electrical)

We now take up the important case of a static metric generated by a mass $M$ with charge $Q$. Assuming spherical symmetry, we obtain the electric field appropriate to a charged mass imbedded in a variable-dielectric-constant medium,

$$\int \vec{D} \cdot \vec{da} = K\varepsilon_o E 4\pi r^2 = Q \tag{63a}$$

or

$$E = \frac{Q}{4\pi K\varepsilon_o r^2} \;. \tag{63b}$$

Substitution into Eq. (59) yields

$$\nabla^2 \sqrt{K} = \frac{1}{\sqrt{K}}\left[\left(\nabla\sqrt{K}\right)^2 - \frac{b^2}{r^4}\right] \tag{64a}$$

or

$$\frac{d^2\sqrt{K}}{dr^2} + \frac{2}{r}\frac{d\sqrt{K}}{dr} = \frac{1}{\sqrt{K}}\left[\left(\frac{d\sqrt{K}}{dr}\right)^2 - \frac{b^2}{r^4}\right], \tag{64b}$$

where $b^2 = Q^2 G/4\pi\varepsilon_o c^4$.

The solution to Eq. (64) as a function of charge (represented by $b$) and mass (represented by $a = GM/c^2$) is given below. Substitution into Eq. (58) verifies that as $r \to \infty$ this expression asymptotically approaches the standard flat-space equations for particle motion about a body of charge $Q$ and mass $M$.

$$\sqrt{K} = \cosh\left(\frac{\sqrt{a^2 - b^2}}{r}\right) + \frac{a}{\sqrt{a^2 - b^2}}\sinh\left(\frac{\sqrt{a^2 - b^2}}{r}\right), \quad a^2 > b^2 \;. \tag{65}$$

(For $b^2 > a^2$ the solution is trigonometric.)

Now let us compare these results for the weak-field case with the comparable Reissner-Nordstrøm solution of the conventional GR approach, which in isotropic coordinates is written [19]

$$ds^2 = g_{ij} dx^i dx^j$$
$$= \left[ \frac{1 - \left(\frac{a}{2r}\right)^2 + \left(\frac{b}{2r}\right)^2}{\left(1 + \frac{a}{2r}\right)^2 - \left(\frac{b}{2r}\right)^2} \right]^2 c^2 dt^2 \quad (66)$$
$$- \left[ \left(1 + \frac{a}{2r}\right)^2 - \left(\frac{b}{2r}\right)^2 \right]^2 (dr^2 + r^2 d\theta^2 + r^2 \sin^2\theta d\phi^2) .$$

Expanding the metric tensor for small departures from flatness as a Maclaurin series in $(a/r) = (GM/rc^2)$, we obtain

$$g_{00} = \left[ \frac{1 - \left(\frac{a}{2r}\right)^2 + \left(\frac{b}{2r}\right)^2}{\left(1 + \frac{a}{2r}\right)^2 - \left(\frac{b}{2r}\right)^2} \right]^2 = 1 - 2\left(\frac{a}{r}\right) + \left(2 + \frac{b^2}{a^2}\right)\left(\frac{a}{r}\right)^2 - \ldots , \quad (67)$$

$$g_{11} = g_{22} = g_{33} = -\left[ \left(1 + \frac{a}{2r}\right)^2 - \left(\frac{b}{2r}\right)^2 \right]^2 = -\left[1 + 2\left(\frac{a}{r}\right) + \ldots\right]. \quad (68)$$

The comparable expansion in the PV approach [see Eqns. (18) - (19)],

$$ds^2 = g_{ij} dx^i dx^j$$
$$= \frac{1}{K} c^2 dt^2 - K(dr^2 + r^2 d\theta^2 + r^2 \sin^2\theta d\phi^2), \quad (69)$$

when expanded by the use of Eq. (65) to the same order as the Reissner-Nordstrøm metric tensor above, yields

$$g_{00} = 1 - 2\left(\frac{a}{r}\right) + \left(2 + \frac{b^2}{a^2}\right)\left(\frac{a}{r}\right)^2 - \ldots , \quad (70)$$

$$g_{11} = g_{22} = g_{33} = -\left[1 + 2\left(\frac{a}{r}\right) + \ldots\right] . \quad (71)$$

Comparison of Eqns. (70) - (71) with Eqns. (67) - (68) reveals that, to the order of expansion shown, the two metric tensors are identical.

We note from Eq. (70) that the sign of the contribution to the metric due to charge (represented by *b*) is counter to that of mass (represented by *a*). In the literature this effect is sometimes discussed in terms of an "electrogravitic repulsion" force [20].

## VI. DISCUSSION

In overview, we have shown that a convenient methodology for investigating general relativistic (GR) effects in a non-abstract formalism is provided by the so-called polarizable-vacuum (PV) representation of GR. The PV approach treats metric perturbation in terms of a vacuum dielectric function $K$ that tracks changes in the effective permittivity and permeability constants of the vacuum, a *metric engineering* approach, so to speak [21]. The structure of the approach is along the lines of the *THεµ* formalism used in comparative studies of gravitational theories. A summary of the important relationships derived here is presented in the accompanying Table.

The PV-derived matter-field Eqns. (58)-(59) are in principle applicable to a wide variety of problems. This short exposition, covering but the Schwarzschild and Reissner-Nordstrøm metrics and the three experimental tests of GR, is therefore clearly not exhaustive. Consideration was confined to cases of spherical symmetry [22], and important topics such as gravitational radiation and frame-dragging effects were not addressed. Therefore, further exploration and extension of the PV approach to specific problems of interest is encouraged, again with cross-referencing of PV-derived results to those obtained by conventional GR techniques to ensure that the PV approach does not generate spurious results.

With regard to the epistemology underlying the polarizable-vacuum (PV) approach as compared with the standard GR approach, one rather unconventional viewpoint is that expressed by Atkinson who carried out a study comparing the two [23]. "It is possible, on the one hand, to postulate that the velocity of light is a universal constant, to define 'natural' clocks and measuring rods as the standards by which space and time are to be judged, and then to discover from measurement that space-time, and space itself, are 'really' non-Euclidean; alternatively, one can *define* space as Euclidean and time as the same everywhere, and discover (from exactly the same measurements) how the velocity of light, and natural clocks, rods, and particle inertias 'really' behave in the neighborhood of large masses. There is just as much (or as little) content for the word 'really' in the one approach as in the other; provided that each is self-consistent, the ultimate appeal is only to convenience and fruitfulness, and even 'convenience' may be largely a matter of personal taste..."

On the other hand, as we have discussed in Section II, from the standpoint of what is actually measured with physical rods and clocks, the conventional approach captures such measurements in a concise, mathematically self-consistent formalism (the tensor approach), despite its somewhat formidable appearance to a neophyte. Therefore, the standard approach is more closely aligned with the positivist viewpoint that underlies modern scientific thought. Nonetheless, the PV model, with its intuitive, physical appeal, can be useful in bridging the gap between flat-space Newtonian physics and the curved-spacetime formalisms of general relativity.

**ACKNOWLEDGEMENTS**

I wish to express my appreciation to G. W. Church, Jr., for encouragement and useful suggestions in the development of this effort.  I also wish to thank M. Ibison and E. Davis for stimulating discussions of the concepts presented herein.

# Table

## Metric Effects in the Polarizable Vacuum (PV) Representation of GR

(For reference frame at infinity, $K = 1$)

| *Variable* | *Determining Equation* | *K≥1* (typical mass distribution, M) |
|---|---|---|
| vel. of light $v_L(K)$ | $v_L = c/K$<br>Eq. (7) | vel. of light $< c$ |
| mass $m(K)$ | $m = m_o K^{3/2}$<br>Eq. (9) | effective mass increases |
| frequency $\omega(K)$ | $\omega = \omega_o / \sqrt{K}$<br>Eq. (10) | redshift toward lower frequencies |
| time interval $\Delta t(K)$ | $\Delta t = \Delta t_o \sqrt{K}$<br>Eq. (11) | clocks run slower |
| energy $E(K)$ | $E = E_o / \sqrt{K}$<br>Eq. (8) | lower energy states |
| length dim. $L(K)$ | $L = L_o / \sqrt{K}$<br>Eq. (13) | objects (spaces) shrink |
| dielectric-vacuum "grav." forces $F_K(K)$ | $F_K(K) \propto \nabla K$<br>Eq. (58) | attractive grav. force |

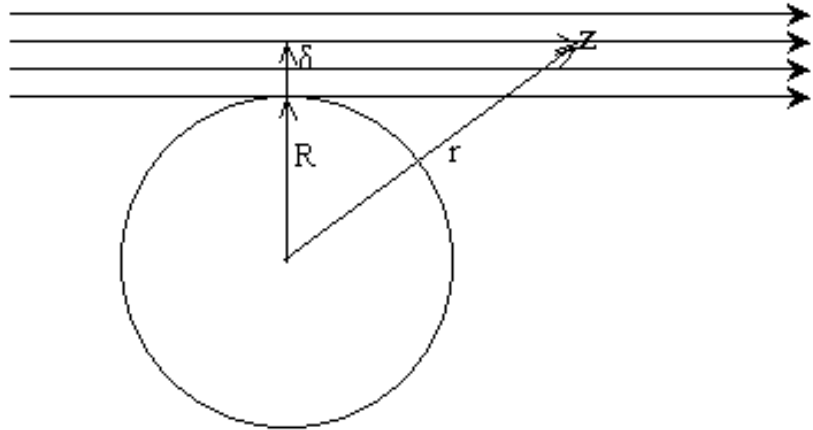

Fig. 1(a)

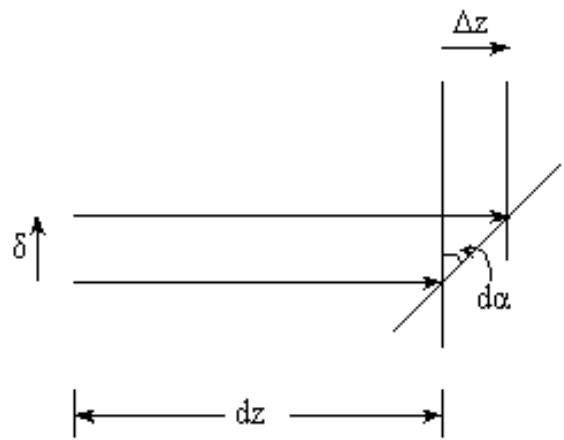

Fig. 1(b)

Fig. 1. Geometry of bending of light rays